\documentclass{ws-p8-50x6-00}
 
\begin{document}
\nopagebreak[1] 
{
\large
\begin{tabbing}
abadabaabadaba   \kill
\end{tabbing} }
 
\vspace{1.0in}
 
{
\large
\begin{tabbing}
\hspace{1.5in} M. A. Garland,\\
\hspace{1.5in} R. E. Schenter,\\
\hspace{1.5in} R. J. Talbert,\\
\hspace{1.5in} S. G. Mashnik, \\
\hspace{1.5in} W. B. Wilson \\
\end{tabbing} }
 
\vspace{-3.6in}
 
{ \Large\bf
\begin{tabbing}
 
\hspace{1.48in} Nuclear Data Requirements for the \\
\hspace{1.5in}  Production Of Medical Isotopes in \\
\hspace{1.5in}  Fission Reactors and Particle Accelerators  \\
\end{tabbing} }
 
\vspace{3.4in}
{
\large
\begin{tabbing}
\hspace{1.5in} 3rd International Conference on Isotopes
\end{tabbing}
\vspace{1.45in}
 
\large
\begin{tabbing}
 
\hspace{1.5in} August 31, 1999
\end{tabbing}
 
\vspace{-6.80in}
 
\hspace{-1.3in}  { \Large\bf LA--UR--99--4898/PNNL--SA--32104\\}
 
\vspace{1.0in}
}
%\newpage
 
%\vspace{.5in}
 
%\hfill {\Large\bf LA--UR--99--draft}

\title{NUCLEAR DATA REQUIREMENTS FOR THE PRODUCTION OF MEDICAL ISOTOPES
IN FISSION REACTORS AND PARTICLE ACCELERATORS }
 
\author{MARC A. GARLAND, ROBERT E. SCHENTER, and ROBERT J. TALBERT}
 
\address{Pacific Northwest National Laboratory, Richland, Washington 99352,USA\\
E-mail: Marc.Garland@pnl.gov, \hspace{.15in} re\_schenter@pnl.gov,
                              \hspace{.15in} bob.talbert@pnl.gov}
 
\author{STEPAN G. MASHNIK and WILLIAM B. WILSON}
 
\address{Los Alamos National Laboratory, Los Alamos, New Mexico 87545, USA \\
         E-mail: mashnik@t2y.lanl.gov, \hspace{.3in} wbw@lanl.gov}
 
\maketitle
 
\vspace{-3.65in}
\begin{tabbing}
\hspace{4.50in} {\Large\bf LA--UR--99--4898}\\
\hspace{4.50in} {\Large\bf PNNL--SA--32104}
\end{tabbing}
\vspace{3.45in}

\abstracts{
       Simulations of isotope-production irradiations of
       targets in the radiation environments of fission
       reactors or particle accelerators require the
       evaluation of proton and/or neutron particle
       fluxes within the target and the integration of
       these with appropriate cross-section data.
       In fission reactors, proton fluxes are typically
       considered negligible and the approach reduces to
       neutron reactions at energies below the upper
       extents of the fission spectrum at which neutrons
       are born ---  10 to 15 MeV.
       In particle accelerator target simulations, the
       secondary neutron flux can provide considerable
       production and/or collateral target activation.
       Where target and product nuclides are separated
       in Z and A, the relative contributions of neutron
       and proton reactions may not be apparent.
       Through decades of effort in nuclear data development
       and simulations of reactor neutronics and accelerator
       transmutation, a collection of reaction data is
       continuing to evolve with the potential of direct
       applications to the production of medical isotopes.
       At Los Alamos the CINDER'90 code and library have been
       developed for nuclide inventory calculations using
       neutron-reaction (En$\leq$20 MeV) and/or decay data for
       3400 nuclides; coupled with the LAHET Code System (LCS),
       irradiations in neutron and proton environments below
       a few GeV are tractable; additional work with the European
       Activation File, the HMS-ALICE code and the reaction
       models of MCNPX (CEM95, BERTINI, or ISABEL with or without 
       preequilibrium, evaporation and fission) have been used to
       produce evaluated reaction data for neutrons and protons to
       1.7 GeV. At the Pacific Northwest National Laboratory, efforts
       have focused on production of medical isotopes and the
       identification of available neutron reaction data from results
       of integral measurements.
           }
 
\section{Introduction}
 
Medical radioisotope production is receiving increased attention due
to the many advances in nuclear medicine. In addition to further
development in diagnostic nuclear medicine, pioneering work is being done
in therapeutic applications of radioisotopes. For example, radiolabeled
monoclonal antibodies are being used to treat leukemia and lymphoma,
brachytherapy is being used to treat prostate cancer, radioactive
stents are being used to prevent restenosis (reclogging of arteries)
following angioplasty treatment of coronary heart disease, and
radioisotopes are being used to palliate the excruciating bone pain
associated with metastatic cancer.
 
Continued success in developing cures for cancer and ultimately in treating
a large number of cancer patients is adversely impacted by this lack of knowledge of
certain neutron capture cross sections for medically important radioisotopes.
Without this data, medical radioisotope production cannot be optimized. Optimization is
not only critical for economic reasons, but also for applications requiring the production
of very high specific activity radioisotopes. In many cases, trial and 
mixed success is ``de rigueur" for producing certain radioisotopes of medical 
significance.
%\vspace{-.55in} 
%\vspace{.35in}

\section{Reactor-Spectrum Data Needs}
 
As an example, the thermal and resonance integral
cross sections are known for $^{186}$W and $^{187}$W , but not for $^{188}$W.
Thus, optimal production of the medically important radioisotope $^{188}$W
may not be realized since calculations related to the design of $^{186}$W
targets, their placements in a reactor and irradiation times cannot accurately be performed.
 
%\vspace{.2in}
%\begin{center}
%\
%$^{187}Re  \rightarrow   ^{188}Re  \rightarrow   ^{189}Re $
%\end{center}
%\vspace{.2in}
 
The design for a reactor focused on isotope production needs to consider the
neutron cross sections of the medical radioisotopes to be produced so that proper
neutronic conditions can be achieved for optimal radioisotope production. Neutron cross
section information is needed to design targets (and hopefully a reactor) for the
optimal production of medical radioisotopes.
 
Production of medical radioisotopes in fission reactor systems must be optimized with
respect to several different parameters: position, target composition, density,
configuration, etc. Research is required to determine the needed cross sections.
Knowledge of these cross sections will benefit several practical applications and will
also provide important modern data information for many isotopes previously
unavailable.
 
The main objective of an initiative to address the cross section deficiencies is to access
the cross sections that are of the greatest projected need. Table~1 identifies
several medical
radioisotopes that harbor deficiencies in cross section knowledge required for efficient,
high specific activity production. In order to demonstrate how the lack of knowledge of
the cross section impacts production results, calculations were made for six important
medical isotope products.
Table 2 shows these results comparing 
values with known and unity cross sections.

\nopagebreak

\begin{table}[t]
\caption{Medical Isotopes with Unknown Cross-Section Data}
\begin{center}
\footnotesize
\begin{tabular}{|c|c|c|c|l|}
\hline
%{} &\raisebox{0pt}[13pt][7pt]{$\Gamma(\pi^- \pi^0)\; s^{-1}$} &
%\raisebox{0pt}[13pt][7pt]{$\Gamma(\pi^-\pi^0\gamma)\; s^{-1}$} &{}\\
\hline
\multicolumn{1}{|c|}{      }    & \multicolumn{1}{c|}{T$_{1/2}$}    &
\multicolumn{1}{c|}{$\sigma_{th}$} & \multicolumn{1}{c|}{R I}    &                                \\
\multicolumn{1}{|c|}{Isotope}   & \multicolumn{1}{c|}{(days)}       &
\multicolumn{1}{c|}{(barns)}    & \multicolumn{1}{c|}{(barns)} & 
\multicolumn{1}{c|}{Medical Application} \\
\hline
\hline
$^{188}W $ &  69.4 &unknown&unknown&cancer and rheumatoid arthritis therapy,  \\
           &       &       &       &radiolabeled antibodies for cancer therapy\\
\hline
$^{186}Re$ &  3.72 &unknown&unknown&prostate cancer and rheumatoid arthritis  \\
           &       &       &       &therapy, radiolabeled antibodies for cancer\\
           &       &       &       &therapy, bone pain palliation             \\
\hline
$^{188}Re$ &  0.71 &unknown&unknown&medullary thyroid carcinoma therapy, bone \\
           &       &       &       &pain palliation, radiolabeled antibodies   \\
           &       &       &       & for cancer therapy                       \\
\hline
$^{194}Os$ &  6.0y &unknown&unknown&radiolabeled antibodies for cancer therapy \\
\hline
$^{193}Os$ &  1.27 &  40   &unknown&cancer therapy                            \\
\hline
$^{198}Au$ &  2.70 & 26E3  &unknown&ovarian, prostate, brain cancer therapy   \\
\hline
$^{166}Ho$ &  1.12 &unknown&unknown&cancer and rheumatoid arthritis therapy   \\
\hline
$^{177}Lu$ &  6.71 & 1000  &unknown&radiolabeled antibodies for cancer therapy,\\
           &       &       &       &heart disease therapy                     \\
\hline
$^{153}Sm$ &  1.93 &  400  &unknown&radiolabeled antibodies for cancer therapy,\\
           &       &       &       &bone pain palliation, treatment of leukemia\\
\hline
$^{153}Gd$ & 241.6 &  2E4  &unknown&osteoporosis detection, SPECT imaging      \\
\hline
$^{127}Xe$ &  36.4 &unknown&unknown&neuroimaging for brain disorders,          \\
           &       &       &       &neuropsychiatric disorder research,       \\
           &       &       &       &SPECT imaging, lung imaging                \\
\hline
$^{125}Xe$ &  0.71 &unknown&unknown&cancer therapy                             \\
\hline
%$^{125}I $ &  59.4 &  900  &  1.4E4&prostate and brain cancer treatment,radio- \\
%           &       &       &       &labeled antibodies for cancer therapy,     \\
%           &       &       &       &osteoporosis detection, SPECT imaging ,    \\
%           &       &       &       &mapping of receptors in the brain          \\
%\hline
$^{126}I $ & 13.0  &unknown&unknown&cancer therapy                             \\

\hline
$^{131}Cs$ & 9.69  &unknown&unknown&intracavity implants for cancer therapy    \\
\hline
\hline
\end{tabular}
\end{center}
\end{table}
\vspace{-.1in}
\begin{table}[t]
\caption{Production Results with Known and Unity Cross Section Values}
\begin{center}
\footnotesize
\begin{tabular}{|c|c|c|c|c|c|c|}
\hline
\hline
\multicolumn{1}{|c|}{            } & \multicolumn{1}{c|}{             } &
\multicolumn{1}{c|}{             } & \multicolumn{1}{c|}{             } &
\multicolumn{1}{c|}{Production   } & \multicolumn{1}{c|}{             } & \multicolumn{1}{c|}{Ratio of  }    \\
 
\multicolumn{1}{|c|}{            } & \multicolumn{1}{c|}{             } &
\multicolumn{1}{c|}{Product      } & \multicolumn{1}{c|}{Target       } &
\multicolumn{1}{c|}{(Ci/g-tgt)} & \multicolumn{1}{c|}{Production   } & \multicolumn{1}{c|}{Unity to  }    \\
 
\multicolumn{1}{|c|}{            } & \multicolumn{1}{c|}{             } &
\multicolumn{1}{c|}{1-Group    } & \multicolumn{1}{c|}{1-Group    } &
\multicolumn{1}{c|}{Using        } & \multicolumn{1}{c|}{(Ci/g-tgt)} & \multicolumn{1}{c|}{Known     }    \\
 
\multicolumn{1}{|c|}{            } & \multicolumn{1}{c|}{             } &
\multicolumn{1}{c|}{Cross        } & \multicolumn{1}{c|}{Cross        } &
\multicolumn{1}{c|}{Known        } & \multicolumn{1}{c|}{Using Unity  } & \multicolumn{1}{c|}{Cross     }    \\
 
\multicolumn{1}{|c|}{Product     } & \multicolumn{1}{c|}{Target       } &
\multicolumn{1}{c|}{Section      } & \multicolumn{1}{c|}{Section      } &
\multicolumn{1}{c|}{Cross        } & \multicolumn{1}{c|}{Cross        } & \multicolumn{1}{c|}{Section   }    \\
 
\multicolumn{1}{|c|}{Isotope     } & \multicolumn{1}{c|}{Isotope      } &
\multicolumn{1}{c|}{(barns)      } & \multicolumn{1}{c|}{(barns)      } &
\multicolumn{1}{c|}{Sections     } & \multicolumn{1}{c|}{Sections     } & \multicolumn{1}{c|}{Production}    \\
\hline
\hline
$^{198}$Au&$^{197}$Au&1.62E+3&1.40E+1& 688&67.4& 0.10 \\
\hline
 $^{60}$Co&$ ^{59}$Co&3.24E-1&5.42E+0&68.1&12.9& 0.19 \\
\hline
$^{125}$I &$^{124}$Xe&1.83E+2&4.61E+1&1190& 151& 0.13 \\
\hline
$^{192}$Ir&$^{191}$Ir&2.89E+1&1.53E+1& 545&47.3&0.087 \\
\hline
$^{145}$Sm&$^{144}$Sm&2.06E+1&1.47E-1&3.91&30.1& 7.7 \\
\hline
$^{153}$Sm&$^{152}$Sm&7.66E+1&3.83E+1&3420&91.5&0.027 \\
\hline
\hline
\end{tabular}
\end{center}
\end{table}
\nopagebreak[4] 
\section{Medium-Energy Data Evaluations}
 
In the radiation environment of a proton accelerator target, neutron
and proton reactions may significantly contribute to the production of
the desired radionuclide. 
Medium-energy protons may each produce a few tens
of neutrons in a high-Z target, each having a significant range and
contribution to particle flux. The complexities resulting from the myriad of
possible reaction paths, along with spatially varying flux magnitudes and 
spectra, require the evaluation of pertinent cross sections and fluxes.
These are evaluated in sequential calculations with the LAHET Code System
LCS\cite{rep89} --- the combination of LAHET and MCNP\cite{Bri93}, or their
subsequent combination in MCNPX\cite{hgh99} --- with the CINDER'90 nuclide
inventory code;\cite{wbw93}$^,$\cite{wbw98} in this sequence, cross sections 
for reactions of protons and medium-energy neutrons are calculated with
on-line nuclear models and evaluated lower-energy neutron reaction cross 
section are contained in the CINDER'90 library. This state-of-the-art 
sequence is used effectively in the analysis of medium-energy designs but
requires a significant investment of CPU time.

Nuclear models have also been utilized with limited available measured
cross section data to form evaluations for a growing number of target 
nuclides. Neutron and proton cross sections from threshold to 1.7 GeV have 
been evaluated for the stable isotopes of O, F, Ne, Na, Mg, Al, 
S, Cl, Ar, K, Zn, Ga, Ge, As, Zr, Nb, Mo, Xe, Cs, Ba, La, and Hg --- or
about 30\%  of the naturally-occuring stable nuclides. These 
evaluations have used available measured data from the LANL T-2 %\cite{ilj94}  
compilation,\cite{ourlib}
the evaluations of the EAF97 library\cite{Sub97} for neutrons 
below 20 MeV, and calculations with HMS-ALICE,\cite{Bla98} 
CEM95,\cite{sgm95} and the BERTINI and ISABEL models of LAHET. Samples
of the data and evaluations for two of nearly 700 reactions evaluated
to date are shown in Figure~1. Complete results are shown in Ref.~10.

\section{Conclusions}

The status of simulation methods and data available for the description of
isotope production is fair and improving, but many additional cross section
measurements and evaluations are needed. Consequently, further research in 
obtaining better cross section information will have positive benefits in
the field of medical science.

%\newpage
 
\begin{figure}[t]
%\figurebox{15}{20pc}{df3b.ps} % to have a box alone
%\epsfxsize=31pc            % will enlarge or reduce the postscript figures based on the xsize
%\epsfxsize=25pc            % will enlarge or reduce the postscript figures based on the xsize
\epsfxsize=20pc          % will enlarge or reduce the postscript figures based on the xsize
\epsfbox{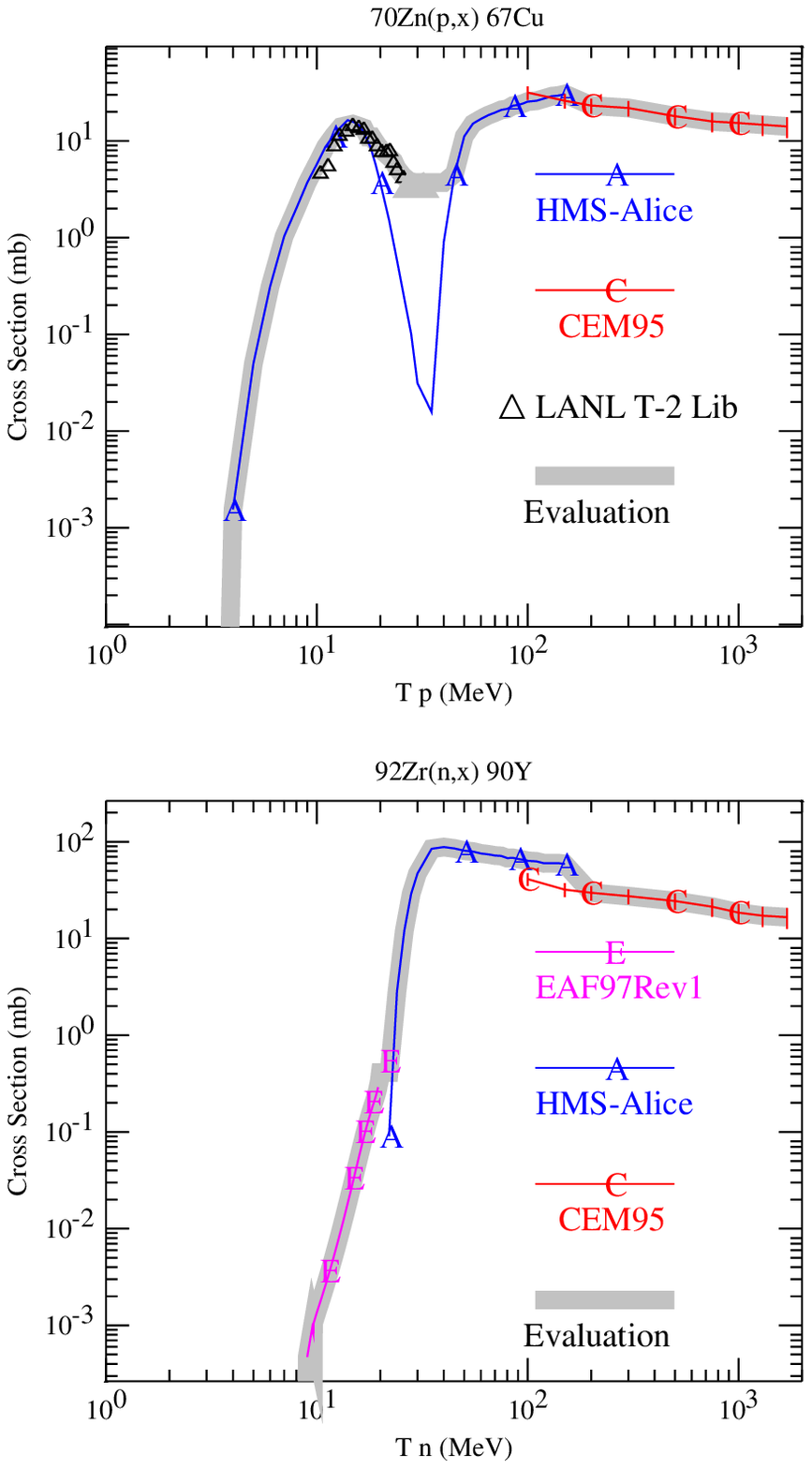}         % postscript image file name
\caption{Samples of Data and Evaluations for (p,x) and (n,x) Reactions.}
\end{figure}

\end{document}